# Magnetization and heat capacity studies of double perovskite compounds $Ba_2SmRuO_6$ and $Ba_2DyRuO_6$

[a,b]Rakesh Kumar[*], [a]C.V. Tomy, [b]R. Nagarajan, [b]P.L. Paulose and [b,c]S.K. Malik

[a]*Indian Institute of Technology Bombay, Powai, Mumbai – 400 076,*
[b]*Tata Institute of Fundamental Research, Colaba, Mumbai – 400 005,*
[c]*International center for condensed matter Physics, University of Brasilia, Brasilia, Brazil*

*Abstract*
*Here we report the magnetic and heat capacity studies of the double perovskite compounds $Ba_2SmRuO_6$ and $Ba_2DyRuO_6$. Antiferromagnetic transitions are inferred at 54 K and 47 K in $Ba_2SmRuO_6$ and $Ba_2DyRuO_6$, respectively in the magnetization measurements. Heat capacity measurements show large jumps at the corresponding temperatures and confirm the bulk magnetic ordering. Both the measurements provide clear indication of the ordering of the rare earth moments also along with the Ruthenium moments. However, the heat capacity results suggest that the ordering of rare earth magnetic moments is spread over a large temperature range and is affected by a large crystal field effect on Ru and rare earth ions. The anomaly observed in the magnetization measurements at 42 K (below the magnetic ordering) in $Ba_2SmRuO_6$ is discerned as a reorientation of $Sm^{3+}$ moments.*

*[*]rakeshk@tifr.res.in*

## 1. Introduction

The double perovskite ruthenates having general formula $M_2Ln$RuO$_6$ ($M$ = alkaline earth metal, $Ln$ = rare earth metal) have been studied due to their interesting structural, magnetic and transport properties [1-9]. These compounds have Ruthenium in the oxidation state of 5+ and show magnetic ordering at low temperatures. In the double perovskites $M_2Ln$RuO$_6$, similar to the perovskite structure $AB$O$_3$, the $A$-sites are occupied by ions of higher co-ordination number and the $B$-sites by ions of relatively lower co-ordination number. Alkaline earth metal ions like Ba or Sr have higher co-ordination number in comparison to that of the rare earth and transition metal ions, and therefore occupy the $A$-site. As a result, the $B$-site is occupied either by the transition or the rare earth metal ions and decides the magnetic properties. In some of the Ba$_2Ln$RuO$_6$ compounds, it is found that [1-3] the ordered Ruthenium moments force the rare earth ions to order at the same temperature, whereas in a few other compounds like $Ba_2ErRuO_6$ [10] and $Ba_2HoRuO_6$ [11], rare earth ions order at relatively lower temperatures than that of the Ruthenium ions. The magnetic properties of many of the Ba$_2Ln$RuO$_6$ compounds have been reported so far, but no such studies exist for the $Ba_2SmRuO_6$ and $Ba_2DyRuO_6$ compounds. Here we present our magnetic and heat capacity measurements on these two compounds and discuss the results.

## 2. Experimental

The compounds $Ba_2SmRuO_6$ and $Ba_2DyRuO_6$ were prepared by the solid state reaction method. Chemicals $BaCO_3$, $Sm_2O_3$, $Dy_2O_3$ and Ru metal powder were used in the preparation of the compounds. Calcination was carried out at 960°C for 24 hours and the final sintering of the pelletized powder was carried out at 1140°C for 24 hours after several intermediate heat treatments followed by grindings. X-ray diffraction (XRD) patterns of the compounds were recorded to check their phase purity. Magnetic properties of these compounds were measured in a SQUID magnetometer (Quantum design, USA). Heat capacity measurements were performed using the relaxation method (PPMS, Quantum design, USA).

### 3. Results and Discussion

*3.1 X-ray diffraction*

XRD patterns of compounds $Ba_2SmRuO_6$ and $Ba_2DyRuO_6$ were recorded at room temperature and could be fitted to a cubic structure having a space group $Fm\bar{3}m$ (No. 225). Lattice parameters and atomic positions were refined by the Rietveld method [12] using Fullprof software, as shown in figure 1. Rietveld analyses of the XRD patterns show the compounds to form in a single phase. The lattice parameters obtained from the analyses are: $a$ = 8.4259(2) Å for $Ba_2SmRuO_6$ and 8.3604(2) Å for $Ba_2DyRuO_6$. Decrease in the lattice parameter from Sm to Dy compound is expected due to the decrease in the ionic radii resulting from the Lanthanide contraction.

*3.2 Magnetic properties*

Temperature dependent magnetic susceptibility for the compounds $Ba_2SmRuO_6$ and $Ba_2DyRuO_6$ were measured in the temperature range of 1.8-300 K. For $Ba_2SmRuO_6$, the magnetic susceptibility plot as a function of temperature, both in the ZFC and FC mode in a magnetic field of 1000 Oe is shown in the main panel of figure 2. Two anomalies are clear, one at 54 K and the second at ~42 K, as shown by arrows in figure 2. No difference between the ZFC and FC susceptibility is observed between 300 K and 54 K where the first anomaly is observed. A bifurcation between the ZFC and FC susceptibility starts below 54 K which becomes significant below 42 K, where the second anomaly is observed, and converges to a Curie-tail below 5 K. Considering the fact that both the Ru and Sm can order in this compound, one can attribute the two anomalies to the magnetic ordering of Ru and Sm. Since the Ru is seen to be ordering first in all the reported compounds of this series, we also assert that the first transition at 54 K is due to the magnetic ordering of the Ru moments. In order to understand the nature of these two anomalies further, the FC magnetization was measured at different field values (see inset of figure 2). It is clear that the FC magnetisation below 42 K is reminiscent of a transition with a ferromagnetic component for small fields, and decreases as the magnetic field is increased. However, the magnetization as a function of magnetic field measured at 5 K, 20 K, 45 K and 60 K gave only a linear variation and does not indicate any ferromagnetic component in the magnetic ordering. Further, no anomaly corresponding to the transition at 42 K is observed in the heat capacity measurements (discussed later). The linear variation of magnetization with magnetic field indicates essentially the antiferromagnetic nature of the magnetic transition. This brings us to the assumption that the magnetic anomaly at 42 K may be due to a reorientation of the magnetic moment [2].

Since Ru is not known to show any type of spin reorientation in this series, we ascertain it to the $Sm^{3+}$ moments, which would have already been ordered along with the Ru moments as in the case of many rare earth moments in this family. It is known that a large exchange field acts at the rare earth site due to the ordered Ru moments which forces the rare earth moments to order at the same temperature as that of the Ru moments in most of the compounds (an exchange field of ~280 kOe is estimated at the Eu site in the isostructural compound $Ba_2EuRuO_6$ from the $^{151}Eu$ Mössbauer measurements [13-14]). Since the magnetic ordering of $Sm^{3+}$ consists of the contributions from the spin as well as the orbital moments, it is possible that one of these components may be getting re-oriented due to the large exchange field acting at the Sm site, as the temperature is decreased below the magnetic ordering. This will also explain the linear behaviour of the magnetization as a function of temperature since the re-orientation does not destroy the net antiferromagnetic ordering in the compound even though magnetization may change slightly with field. As the field value is increased, this reorientation also gets suppressed, decreasing the value of the susceptibility as observed (inset of figure 2).

In order to understand the contributions of Ru and Sm ions to magnetism of the material, an analysis of their contribution to the paramagnetic state was carried out. It is well known that in the case of $Sm^{3+}$ ions, there is a mixing of magnetic properties of the excited states with that of its ground state. Therefore, the paramagnetic susceptibility of free $Sm^{3+}$ is not a simple curie weiss law, but is a complicated function of temperature (discussed latter). Even though, the total susceptibility was found to obey the Curie Weiss behaviour from 300 K down to the magnetic ordering temperature (shown as the solid line in figure 3). However, the effective paramagnetic moment value obtained from this fit ($\sim 5\mu_B$) is very unrealistic to be arising from the combined paramagnetic susceptibility of Ru and Sm ions ($p_{eff} = \sqrt{\mu_{Sm^{3+}}^2 + \mu_{Ru^{5+}}^2} = \sqrt{(1.5)^2 + (3.87)^2} = 4.2 \mu_B$). The reason for such a large value is not clear at present. We have tried to obtain the contributions of Ru and Sm ions independently to magnetism of the material. If the crystal field effects are neglected, then the paramagnetic susceptibility of $Sm^{3+}$ can be calculated using the standard expression (which includes contributions from the higher energy states in an angular momentum multiplet $J$ [15])

$$\chi_{mol}(Sm^{3+}) = \frac{0.1241}{xT}\left[\frac{2.14x + 3.67 + (42.9x + 0.82)e^{-7x} + (142x - 0.33)e^{-16x} + .....}{3 + 4e^{-7x} + 5e^{-16x} + 6e^{-27x} + ....}\right]$$

Here $T$ is temperature and the value of $x$ depends upon the screening constant $\sigma$. We attempted the computation with values of $x = 220/T$ and $191/T$ corresponding to σ = 33 and σ = 34, respectively [15]. We found that $x = 191/T$ describes our observations of $Ba_2SmRuO_6$ better. The result of this calculation of the paramagnetic susceptibility of $Sm^{3+}$ is plotted in figure 4 along with the observed total magnetic susceptibility of $Ba_2SmRuO_6$. The paramagnetic susceptibility of free $Sm^{3+}$ ions overshoots the magnetic susceptibility value of $Ba_2SmRuO_6$ at 22 K, which suggests that $Sm^{3+}$ ions would have ordered antiferromagnetically at a temperature higher than 22 K. Considering the proper-ties reported for analogous systems [13-14], we believe that the magnetic order of Sm magnetic moments are induced by the ordering of Ru magnetic moments. In order to obtain the paramagnetic moment of $Ru^{5+}$ ions, the calculated Sm susceptibility contribution was subtracted from the total observed susceptibility and the resultant inverse susceptibility was fitted to the Curie-Weiss law which is shown in the inset of figure 4. It is clear that the Curie-Weiss fit deviates from the straight line behaviour at relatively high temperatures, *i.e.*, below ~220 K. This may be due to the crystal field effects experienced by Ru ions, which is established in other compounds in this series [1]. The obtained value of the magnetic moment per Ru ion, 3.88(1) $\mu_B$, is also close to the expected value of 3.87 $\mu_B$ when the orbital magnetic moment is totally quenched due to the crystal field effects. It is to be kept in mind that in the above consideration, effect of crystal field on $Sm^{3+}$ ions was neglected, whereas, the crystal field effects on Sm ions could be significant.

Magnetic susceptibility as a function of temperature for $Ba_2DyRuO_6$ is shown in figure 5. The magnetic susceptibility increases as the temperature is decreased below 300 K. A small anomaly is observed at low temperatures, which manifests itself as a change in the slope of magnetic susceptibility at ~47 K, shown by the arrow (a clear peak is observed in the heat capacity of this compound at this temperature). On the basis of the general trend of Ruthenium ordering first in the double perovskite compounds, the 47 K anomaly can be assigned to the magnetic ordering of the Ru moments. However, the magnetic susceptibility behaves like a paramagnetic material down to 1.8 K except for a small change in slope at the above mentioned temperature. Hence it is not very clear whether the Dy moments also order in this compound. In order to investigate further, we simulated the paramagnetic susceptibility of $Dy^{3+}$ ions using $\theta_p = -17.8(4)$ K (Table 1) is shown in figure 5. Since the paramagnetic magnetic moment of $Dy^{3+}$ (10.6 $\mu_B$) [16] ions is far greater than that of the $Ru^{5+}$ (3.87 $\mu_B$) ions, the resultant paramagnetic moment ($p_{eff} = \sqrt{\mu_{Dy^{3+}}^2 + \mu_{Ru^{5+}}^2}$ = 11.28 $\mu_B$), and hence the resultant paramagnetic susceptibility should be close to that of the Dy ions. The paramagnetic susceptibility of free $Dy^{3+}$ ions overshoots the observed magnetic susceptibility of $Ba_2DyRuO_6$ at 38 K, and becomes very prominent below it. If only the $Ru^{5+}$ moments order, then the paramagnetic susceptibility of $Ba_2DyRuO_6$ should have followed the simulated paramagnetic susceptibility of $Dy^{3+}$ below 38 K. This suggests that the $Dy^{3+}$ moments also order at ~47 K, with the nature of order being dominantly antiferromagnetic (since no anomaly is observed in heat capacity except for 47 K). The continued increase in susceptibility below the magnetic ordering temperature may be due the presence of disorder as indicated in the case of the Sm compound above (see also heat capacity results below), except that in this case a considerable fraction of Dy ions continued to be in the paramagnetic state even at 1.8 K. It may be noted here that in another compound of this family, $Ba_2EuRuO_6$, the presence of disorder and paramagnetic component down to 1.8 K were inferred from the development of $^{151}Eu$ Mössbauer hyperfine fields [13-14]. Magnetization as a function of magnetic field at 5 K, 10 K, 30 K, 40 K and 50 K varies linearly with magnetic field (not shown here), which suggests the antiferromagnetic nature of the magnetic interactions.

The inverse magnetic susceptibility of $Ba_2DyRuO_6$ is fitted to the Curie-Weiss law in the temperature range of 50-300 K (inset of figure 5). The parameters obtained from the Cu-rie-Weiss fit are $\mu_{eff}$ = 11.08(1) $\mu_B$ and $\theta_p$ = −17.8(4) K. The effective magnetic moment is close to the expected value (11.28 $\mu_B$). If we take the magnetic moment for free $Dy^{3+}$ ions to be 10.6 $\mu_B$ [16], then the magnetic moment of $Ru^{5+}$ in $Ba_2DyRuO_6$ is estimated to be 3.24(1) $\mu_B$. The effective magnetic moment of $Ru^{5+}$ is calculated (theoretically) to be 0.77 $\mu_B$ on lifting the degeneracy due to spin-orbit coupling and can increase to 3.87 $\mu_B$ on quenching of the orbital magnetic moment due to crystal field effects. The estimated effective moment of $Ru^{5+}$ in $Ba_2DyRuO_6$ (3.24 $\mu_B$) lies between the above two values and suggests a large crystal field effect on $Ru^{5+}$ ions. The lower value of $Ru^{5+}$ moments may also be due to the crystal field effect on $Dy^{3+}$ moments, which we have neglected in simulating the paramagnetic susceptibility of $Dy^{3+}$ moments. The negative paramagnetic Curie temperature (Table 1) obtained from the Curie-Weiss fit to the inverse susceptibility indicates the presence of antiferromagnetic interactions in these compounds.

*3.3 Heat capacity measurements*

Heat capacity (C) measurements on the compounds $Ba_2SmRuO_6$ and $Ba_2DyRuO_6$ were performed in a temperature range of 1.8-100 K. The measured heat capacity for $Ba_2SmRuO_6$ is shown in the lower panel of figure 6 which shows a clear jump of ~12 Jmol$^{-1}$K$^{-1}$ at the magnetic ordering temperature of ~54 K confirming the bulk nature of magnetic order in this material. In order to separate the magnetic contribution, heat capacity of isostructural and isomorphous nonmagnetic analog compound $Ba_2LuNbO_6$ was subtracted from the total heat capacity [10]. This magnetic contribution to the heat capacity is shown in the upper panel of figure 6 along with the calculated magnetic entropy $S_{mag}(=\int_{T_1}^{T_2}\frac{C_{mag}}{T}dT)$ as a function of temperature. Magnetic entropy increases with temperature and nearly saturates above 54 K with a value of ~14 Jmol$^{-1}$K$^{-1}$ at 60 K. In double perovskites, the magnetic transition is primarily due to the ordering of $Ru^{5+}$ ions which have a ground state of $J$ = 3/2 corresponding to four degenerate states +3/2, +1/2, −1/2 and −3/2. In the presence of crystalline electric field, the four degenerate levels would split into two levels, $|J=3/2, M_J=\pm 3/2>$ and $|J=3/2, M_J=\pm 1/2>$ each consisting of two degenerate levels giving rise to a multiplicity of only two [10-11]. Hence, the magnetic entropy of the compound should be $R\ln(2\times\frac{1}{2}+1)=5.76$ Jmol$^{-1}$K$^{-1}$, rather than $R\ln(2\times\frac{3}{2}+1)=11.52$ Jmol$^{-1}$K$^{-1}$. This implies that the observed magnetic entropy ~14 Jmol$^{-1}$K$^{-1}$ should include other magnetic contributions also. The observed additional contribution of ~8.24 Jmol$^{-1}$K$^{-1}$ can arise if the $Sm^{3+}$ ions (which has a ground state $J$ = 5/2) also order magnetically. However, if all the $Sm^{3+}$ ions in $Ba_2SmRuO_6$ would have ordered then the magnetic entropy should have been $R\ln(2\times\frac{5}{2}+1)=14.9$ Jmol$^{-1}$K$^{-1}$. If we assume the full contribution from Ru ions (as in the case of nonmagnetic rare earth compounds [10]), this discrepancy in the magnetic entropy implies that only a fraction (~55%) of $Sm^{3+}$ ions order in this compound. If we consider only the partial ordering of the $Sm^{3+}$ ions, then the unordered $Sm^{3+}$ ions should have given a large contribution to the paramagnetic susceptibility at low temperatures, which is not observed in the experimental data (see figure 3). In order to verify whether the discrepancy in the magnetic entropy is arising due to the crystal field effects, we consider the following. $Sm^{3+}$ has a ground state of $J = 5/2$, giving rise to the six degenerate levels. In the presence of crystal field effects, the degeneracy will be removed and the six-fold degenerate ground state will be split into three distinct energy doublets, $\pm 5/2$, $\pm 3/2$ and $\pm 1/2$, giving a magnetic entropy of $R\ln 3 = 9.14\,\text{Jmol}^{-1}\text{K}^{-1}$ which is comparable with the observed value of 8.24 $\text{Jmol}^{-1}\text{K}^{-1}$. Thus, the above considerations suggest that in this compound, both the $Ru^{5+}$ ions and $Sm^{3+}$ ions order magnetically. It is also to be noted that the heat capacity below the magnetic ordering shows a broad hump instead of a sharp λ-type anomaly, as seen in the non magnetic rare earth analogue $Ba_2YRuO_6$ [10]. This indicates that all the Sm ions do not order at once but the order is spread over a large temperature range, possibly resulting in a magnetic disorder [14-15]. The change in slope of the entropy curve around ~42 K may indicate the reorientation of the magnetic moments, which was seen as the magnetic anomaly in the susceptibility around 42 K (but not seen in the heat capacity) [2].

Heat capacity of $Ba_2DyRuO_6$ is shown in the lower panel of figure 7 along with its isostructural nonmagnetic analog $Ba_2LuNbO_6$ [10]. The behaviour of the heat capacity of the Dy-compound is similar to that of the Sm-compound (figure 6). The jump in heat capacity observed at ~47 K confirms the suggestion of magnetic transition inferred from the anomaly seen in magnetic susceptibility around the same temperature. The extent of the jump (~15 $\text{Jmol}^{-1}\text{K}^{-1}$) also confirms the bulk nature of the magnetic order. Magnetic contribution to the heat capacity ($C_{mag}$) and the corresponding change in entropy ($S_{mag}$) was estimated in the same way as discussed above in the case of the Sm compound. Magnetic entropy increases with temperature and becomes nearly constant above 50 K to a value of 15.62 $\text{Jmol}^{-1}\text{K}^{-1}$. Following the same arguments as presented above in the case of Sm compound, it is inferred that the magnetic entropy includes contributions from both the $Ru^{5+}$ ions and from the $Dy^{3+}$ (ground state $J = 15/2$) ions. A complete ordering of the Dy ions would have contributed 23.05 $\text{Jmol}^{-1}\text{K}^{-1}$ to the magnetic entropy in addition to 5.76 $\text{Jmol}^{-1}\text{K}^{-1}$ from the ordering of $Ru^{5+}$ ions. Hence the extra contribution to the magnetic entropy from the $Dy^{3+}$ moments corresponds to only 42.4% of the ordered Dy ions. $Dy^{3+}$ has a ground state of $J = 15/2$. In the presence of crystal fields, the sixteen states of the $Dy^{3+}$ ions would split into eight degenerate doublets which would give a magnetic entropy contribution of 17.28 $\text{Jmol}^{-1}\text{K}^{-1}$ if all the $Dy^{3+}$ moments order. However, the additional contribution is only 9.86 $\text{Jmol}^{-1}\text{K}^{-1}$. This suggests all Dy moments might not have ordered and that there is a large disorder, as suggested above in the case of the Sm compound. This is also confirmed by the fact that the heat capacity below the ordering temperature has a broad hump, suggesting spread of magnetic ordering. This in turn also explains the increase in the magnetic susceptibility below the magnetic ordering temperature.

### 4. Conclusions

From the magnetic and heat capacity studies of $Ba_2SmRuO_6$ and $Ba_2DyRuO_6$ we infer that $Ru^{5+}$ ions order antiferromagnetically in these compounds at ~54 K and 47 K, respectively. The results also suggest that the rare earth moments order along with the Ru moments. However, the features of the magnetic susceptibility and heat capacity below the ordering temperature indicate that there is considerable magnetic disorder, which we suggest is mostly confined to the rare earth moments, though the magnetic disorder amongst Ru moments can not be ruled out. Crystal field effects are also felt in both the compounds. Detailed neutron diffraction measurements are needed to ascertain the exact nature of magnetic orderings in these compounds.

**Reference:**

1. Izumiyama Y, Doi Y, Wakeshima M, Hinatsu Y, Shimojo Y and Morii Y 2001 *J. Phys.: Condens. Matter* **13** 1303

2. Izumiyama Y, Doi Y, Wakeshima M, Hinatsu Y, Oikawa K, Shimojo Y and Morii Y 2000 *J. Mater. Chem.* **10** 2364

3. Doi Y, Hinatsu Y, Nakamura A, Ishii Y and Morii Y 2003 *J. Mater. Chem.* **13** 1758

4. Doi Y and Hinatsu Y 1999 *J. Phys.: Condens. Matter* **11** 4813

5. Parkinson N G, Hatton P D, Judith A K H, Ritter C, Ibberson R M and Wu M K 2004 *J. Phys.: Condens. Matter* **16** 7611

6. Doi Y, Hinatsu Y, Oikawa K, Shimojo Y and Morii Y 2000 *J. Mater. Chem.* **10** 797

7. Greatrex R, Greenwood N N, Lal M and Frenandez I 1979 *J. Solid State Chem.* **30** 137

8. Battle P D, Jones C W and Studer F 1991 *J. Solid State Chem.* **90** 302

9. Battle P D, Goodenough J B and Price R 1983 *J. Solid State Chem.* **46** 234

10. Izumiyama Y, Doi Y, Wakeshima M, Hinatsu Y, Nakamura A and Ishii Y 2002 *J. Solid State Chem.* **169** 125

11. Hinatsu Y, Izumiyama, Doi Y, Alemi A, Wakeshima M, Nakamura A and Morii Y 2004 *J. Solid State Chem.* **177** 38

12. Rodriguez-Carvajal Juan 2001 *An Introduction to the program Fullprof 2000* (Version July 2001)

13. Gibb T C and Greatrex R 1980 *J. Solid State Chem.* **34** 279

14. Kumar R, Tomy C V, Paulose P L and Nagarajan R 2005 *J. Appl. Phys.* **97** 10A907

15. Frank A 1932 *Phys. Rev.* **39** 119

16. Van Vleck J H, *The Theory of Electric and Magnetic Susceptiblities,* Oxford 1952, p.243

**Figure Captions:**

**Figure 1.** Rietveld analyses of the XRD patterns of the compounds $Ba_2SmRuO_6$ and $Ba_2DyRuO_6$.

**Figure 2.** ZFC and FC magnetic susceptibility as a function of temperature for $Ba_2SmRuO_6$ in a magnetic field of 1000 Oe. Inset shows the plot of FC magnetic susceptibility near the transition temperature at various applied fields.

**Figure 3.** Measured magnetic susceptibility for $Ba_2SmRuO_6$ (closed circle) in a magnetic field of 5000 Oe. Open circle shows the inverse susceptibility fitted to Curie-Weiss law in the temperature range of 50-300 K.

**Figure 4.** Measured magnetic susceptibility for $Ba_2SmRuO_6$ (closed circle), calculated paramagnetic susceptibility for $Sm^{3+}$ (open triangle) and the difference between the two (closed triangle), as a function of temperature. Inset shows the inverse difference susceptibility fitted to the Curie-Weiss law.

**Figure 5.** Measured magnetic susceptibility for $Ba_2DyRuO_6$ (open circle) and the calculated magnetic susceptibility for $Dy^{3+}$ ions (closed triangle), as a function of temperature. Inset shows the inverse susceptibility fitted to the Curie-Weiss law in the temperature range 50-300 K.

**Figure 6.** Lower panel: Heat capacity of $Ba_2SmRuO_6$ (closed square) and its nonmagnetic analog compound $Ba_2LuNbO_6$ (line) [10]. Upper panel: Magnetic heat capacity and the corresponding change in magnetic entropy.

**Figure 7.** Lower panel: Heat capacity of $Ba_2DyRuO_6$ (closed square) and its nonmagnetic analog compound $Ba_2LuNbO_6$ (line) [10]. Upper panel: Magnetic heat capacity and the corresponding magnetic entropy as a function of temperature for $Ba_2DyRuO_6$.

**Table 1.** Neel temperature ($T_N$), magnetic moment ($\mu_{Ru5+}$) estimated for $Ru^{5+}$ ions and paramagnetic Curie temperature ($\theta_p$) from the Curie-Weiss fit of the paramagnetic susceptibility.

| Compound | $T_N$ (K) | $\mu_{Ru5+}$ ($\mu_B$) | $\theta_p$ (K) |
|---|---|---|---|
| $Ba_2SmRuO_6$ | 54 | 3.88(1) | − 447(4) |
| $Ba_2DyRuO_6$ | 47 | 3.24(1) | − 17.8(4) |

**Figure 1**

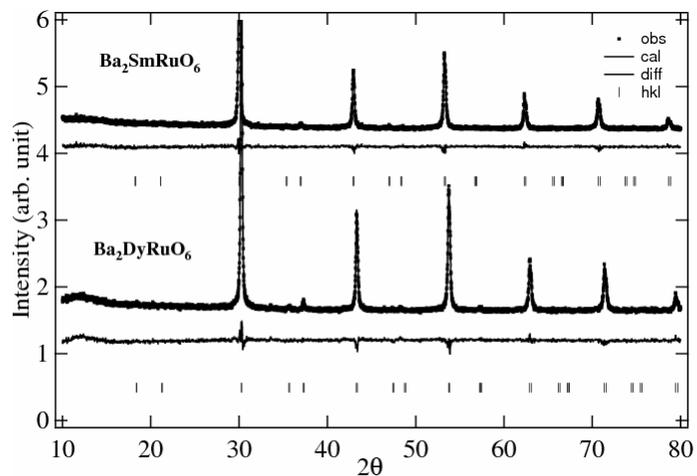

**Figure 2**

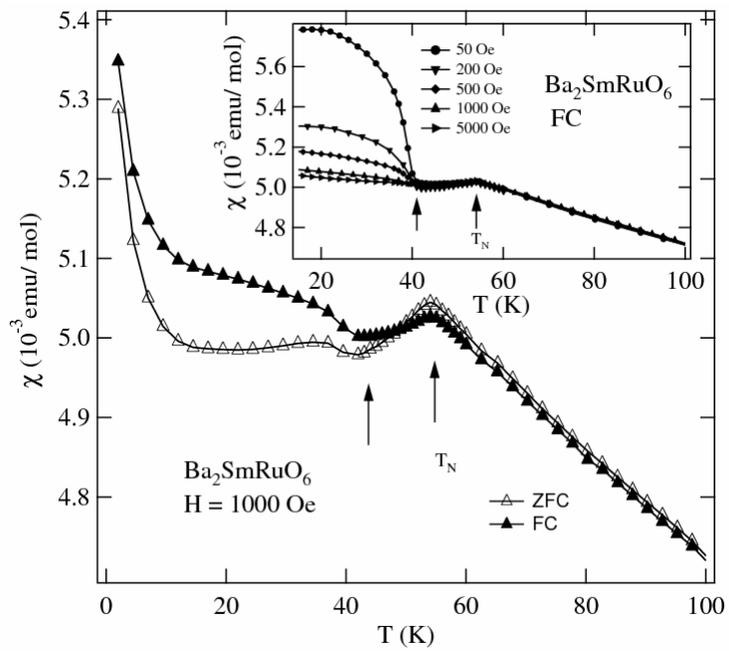

Figure 3

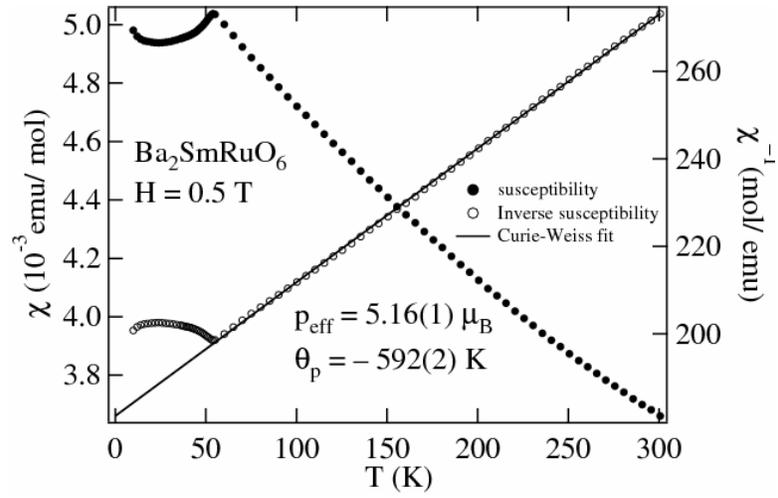

Figure 4

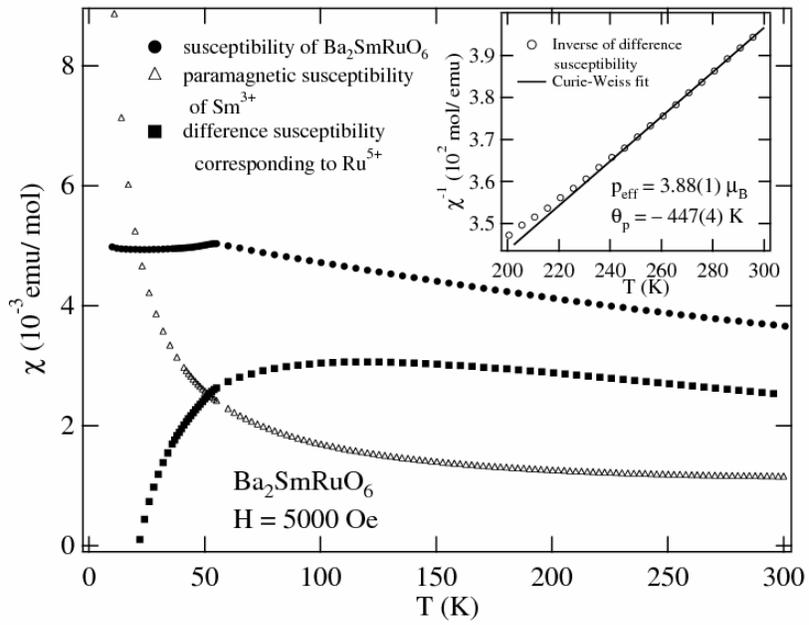

**Figure 5**

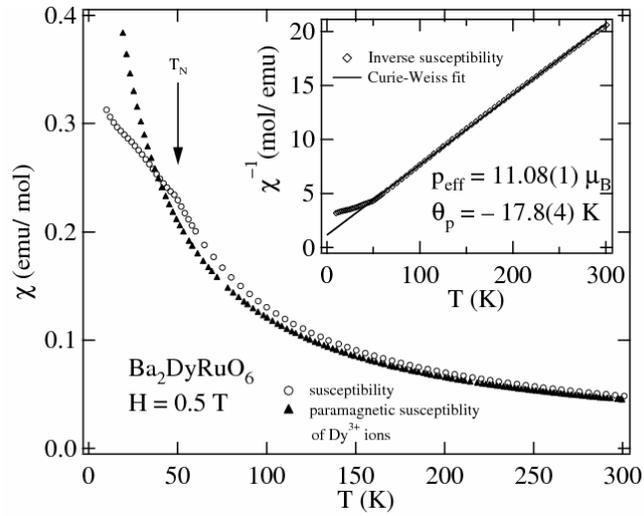

**Figure 6**

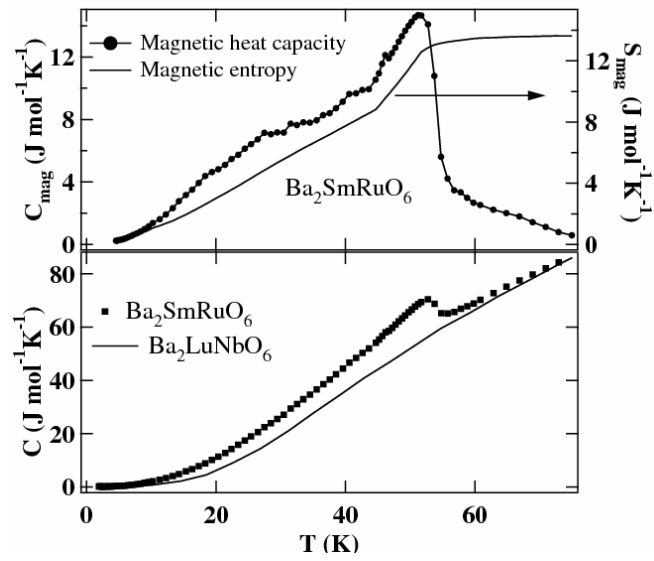

**Figure 7**

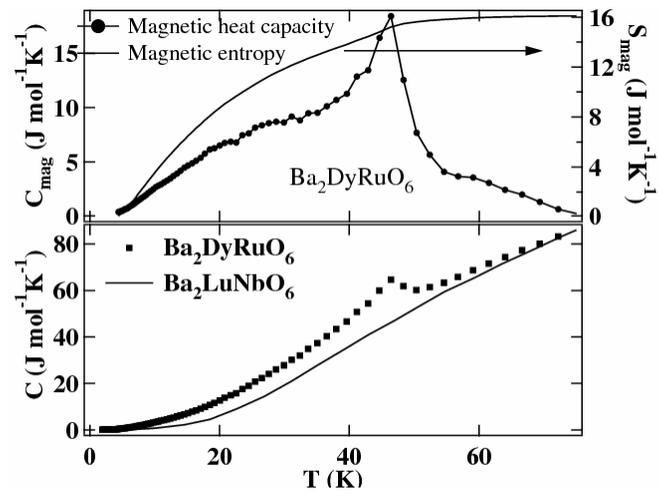